\documentclass[aps,showkeys,twocolumn,preprintnumbers,amsmath,amssymb,superscriptaddress,floatfix,nofootinbib]{revtex4-2}

\usepackage{graphicx,color,booktabs,bm}
\usepackage{amsmath}
\usepackage{slashed}
\usepackage{txfonts}
\usepackage{overpic}
\usepackage{epsfig}
\usepackage{amssymb}
\usepackage{bbding}
\usepackage{epstopdf}
\usepackage{appendix}
\usepackage{indentfirst}
\usepackage{feynmf}
\usepackage{cases}
\usepackage{multirow}

\graphicspath{{Figures/}}

\usepackage[colorlinks=true, citecolor=blue, linkcolor=red, urlcolor=blue]{hyperref}
\begin{document}
\title{Probing the internal structure of the \(\Omega\Omega\) dibaryon via its weak decays and lifetime}

\author{Qi-Fang L\"{u}}\email{lvqifang@hunnu.edu.cn}

\affiliation{Department of Physics, Hunan Normal University, Changsha 410081, China}
\affiliation{Key Laboratory of Low-Dimensional Quantum Structures and Quantum Control of Ministry of Education, Changsha 410081, China}
\affiliation{Key Laboratory for Matter Microstructure and Function of Hunan Province, and Hunan Research Center of the Basic Discipline for Quantum Effects and Quantum Technologies, Hunan Normal University, Changsha 410081, China}

\author{Deng-Wang Tang}
\email{dengwangtang@hunnu.edu.cn}
\affiliation{Department of Physics, Hunan Normal University, Changsha 410081, China}

\author{Yu-Bing Dong}
\email{dongyb@ihep.ac.cn}
\affiliation{Institute of High Energy Physics, Chinese Academy of Sciences, Beijing 100049, China}
\affiliation{School of Physical Sciences, University of Chinese Academy of Sciences, Beijing 100049, China}

\author{Peng-Nian Shen}
\email{shenpn@ihep.ac.cn}
\affiliation{Institute of High Energy Physics, Chinese Academy of Sciences, Beijing 100049, China}

\author{Jia-Jun Wu}
\email{wujiajun@ucas.ac.cn}
\affiliation{School of Physical Sciences, University of Chinese Academy of Sciences, Beijing 100049, China}
\affiliation{Southern Center for Nuclear-Science Theory (SCNT), Institute of Modern Physics, Chinese Academy of Sciences, Huizhou 516000, China}

\begin{abstract}
	The weak decay properties of the $\Omega\Omega$ dibaryon play a crucial role in probing its internal structure and guiding experimental efforts. Using an effective Lagrangian approach, we systematically study its weak decays and lifetime by computing the widths for both three- and four-body final states, with amplitudes receiving contributions from single- and double-weak-vertex diagrams. We investigate the dependence of the decay properties on the binding energy ranging from a few MeV to over 100 MeV. For weakly bound states ($E_b = 1$ and $20$ MeV), the three-body channels $\Omega \Lambda \bar{K}$, $\Omega \Xi \pi$, and $\Xi \Xi \bar{K}$ exhaust essentially the entire width, with the largest channel shifting from $\Omega \Lambda \bar{K}$ at $E_b=1$ MeV to $\Omega \Xi \pi$ at $E_b=20$ MeV. For deeply bound states ($E_b = 100$ and $120$ MeV), the $\Omega \Lambda \bar{K}$ channel is forbidden, and the dominant modes are $\Omega \Xi \pi$ and $\Xi \Xi \bar{K}$, while the double-weak-vertex channel $\Lambda \bar{K} \Lambda \bar{K}$ is found to be negligibly small. As the binding energy increases, the total lifetime extends from the order of $10^{-10}~\mathrm{s}$ to $10^{-9}~\mathrm{s}$. These distinct differences in decay patterns and lifetimes provide valuable theoretical reference values for experimental searches at RHIC and the LHC, and offer a clear discriminator between the compact hexaquark and weakly bound molecular interpretations.
\end{abstract}


\maketitle

\section{Introduction}\label{introduction}

The study of multiquark states beyond the conventional meson and baryon configurations has been one of the central theme in hadron physics~\cite{Guo:2017jvc,Dong:2017gaw,Liu:2019zoy,Brambilla:2019esw,Guo:2019twa,Chen:2022asf,Meng:2022ozq,Mai:2022eur,Liu:2024uxn,Garcilazo:2025wkt,Wang:2025sic}. Among these exotic configurations, dibaryons with baryon number $B=2$ occupy a special position because they directly probe the interplay between quark confinement and baryon-baryon interactions~\cite{Clement:2016vnl,Clement:2020mab,Dong:2023xdi,Bashkanov:2023yca,Dai:2023ofz}. The historical trajectory of dibaryon research reveals a clear progression, from systems composed only of light $u$ and $d$ quarks to those with increasing strangeness content, and even to systems containing heavy flavor quarks. The deuteron, the only confirmed stable dibaryon, is a weakly bound $pn$ system with a binding energy of merely 2.22 MeV. The $d^*(2380)$ resonance observed by WASA-at-COSY collaboration~\cite{WASA-at-COSY:2011bjg,WASA-at-COSY:2014dmv,WASA-at-COSY:2014lmt}, has been usually interpreted as a deeply bound hexaquark compared with the $\Delta\Delta$ threshold~\cite{Dyson:1964xwa,Yuan:1999pg,Li:1999dm,Ping:2000dx,Dai:2005kt,Bashkanov:2013cla,Gal:2013dca,Huang:2013nba,Chen:2014vha,Park:2015nha,Huang:2015nja,Lu:2017uey,Lu:2024dtb}, although its exact nature remains a subject of debate~\cite{Ikeno:2021frl,Molina:2021bwp}. The introduction of strangeness, however, fundamentally alters the theoretical landscape. Strangeness enriches the internal structure of dibaryons, modifies the short-range QCD dynamics through one-gluon exchange, light-meson exchange, and the Pauli principle, and opens a rich variety of new decay channels that are absent in non-strange systems. Crucially, the mass spectrum of a dibaryon determines whether such a state exists, while the decay properties determine whether it can be observed and how its internal structure can be revealed. A state that cannot decay strongly but decays only weakly acquires a relatively long lifetime, providing a distinctive experimental signature and making it a promising target for experimental searches. This is precisely the case for the $\Omega \Omega$ dibaryon, the most strange dibaryon with strangeness $S=-6$, which provides a particularly clean laboratory for studying weak decays in a multiquark environment.

The significance of the \(\Omega\Omega\) dibaryon extends well beyond exotic hadron spectroscopy, with important implications for nuclear astrophysical systems. In the dense interior of neutron stars, where densities can reach several times the nuclear saturation density, the appearance of hyperons is energetically favored and is expected to soften the equation of state, which in turn lowers the predicted maximum neutron-star mass~\cite{Tolos:2020aln}. This tension between the predicted softening and the observation of neutron stars with masses exceeding about two solar masses is known as the hyperon puzzle and remains a central problem in nuclear astrophysics~\cite{Haidenbauer:2016vfq,Yuan:2026lum,Antoniadis:2013pzd,Glendenning:1998wp,NANOGrav:2019jur}. The possible formation of dibaryonic states, such as the \(H\) dibaryon~\cite{Glendenning:1998wp}, \(d^*(2380)\)~\cite{Vidana:2017qey} or even \(\Omega\Omega\) dibaryon, introduces a new degree of freedom that could further modify the equation of state and the thermal evolution of neutron stars. The \(\Omega \Omega\) interaction also provides a clean probe of the poorly constrained strangeness \(S=-6\) sector of hyperon-hyperon interactions, which are difficult to access in terrestrial experiments. Moreover, the weak decay properties of the \(\Omega \Omega\) dibaryon are essential for determining whether such a state can survive long enough inside a neutron star to influence its macroscopic properties, or whether it would decay before affecting the equation of state. Therefore, while current astrophysical observations, including neutrino emission signals from supernova SN1987A~\cite{McDermott:2018ofd}, neutron-star mass measurements~\cite{Mantziris:2020xwi}, and dark-matter direct detection~\cite{Kolb:2018bxv}, can only provide indirect limits on various exotic dibaryon states, a direct experimental confirmation of the \(\Omega\Omega\) system in the future would be a crucial signal for resolving its fundamental properties and addressing broader issues such as neutron star Bose-Einstein condensation and heavy-ion femtoscopy. These connections make a quantitative understanding of the \(\Omega\Omega\) system, including its decay behavior, a problem of broad interest across hadron physics and nuclear astrophysics.

Despite these compelling motivations from both hadron physics and nuclear astrophysics, the fundamental properties of the $\Omega\Omega$ dibaryon, such as its binding energy and internal structure, remain far from settled. Predictions for its binding energy span an enormous range~\cite{Zhang:2000sv,Zhang:2000mm,Li:2000cb,Zhang:2003fk,Dai:2006gs,Pang:2002ga,Huang:2019hmq,Gongyo:2017fjb,Dhindsa:2025gae,Liu:2002vi,HALQCD:2015qmg,Chen:2019vdh,Deng:2023zlx,Wan:2025zau,You:2025ksc}, from deeply bound states of order 100 MeV to weakly bound systems near threshold, and even to unbound configurations in certain channels. With the chiral SU(3) quark model and its extended version, by which the most experimental data on the baryon-baryon scatterings, reactions and binding behaviors are well reproduced, the $\Omega \Omega$ state with $J^P=0^+$ was predicted to be deeply bound, with a binding energy over 100 MeV and a root-mean-square distance with 0.84 fm~\cite{Zhang:2000sv,Zhang:2000mm,Li:2000cb,Zhang:2003fk,Dai:2006gs}. With the extended quark-delocalization color-screening model, the $\Omega\Omega$ dibaryon was found to lie about $0.3\text{--}92.3~\mathrm{MeV}$ below the $\Omega\Omega$ threshold~\cite{Pang:2002ga,Huang:2019hmq}. In sharp contrast to these quark-model predictions, the Lattice QCD calculations from the HAL QCD Collaboration, performed with a nearly physical pion mass of \(m_\pi=146~{\rm MeV}\) and a large volume of \((8.1~{\rm fm})^3\), yield a weakly bound \(\Omega\Omega\) state with a binding energy of \(1.6(6)^{+0.7}_{-0.6}~{\rm MeV}\) in the \(^1S_0\) channel~\cite{Gongyo:2017fjb}. In addition, more recent Lattice QCD studies have found that the \(\Omega\Omega\) system in the \(^1S_0\) channel lies very close to threshold, making definitive conclusions difficult, and that it is unbound in the \(^5S_2\) channel~\cite{Dhindsa:2025gae}. This disagreement across frameworks, from deeply bound quark-model states to weakly bound or even unbound Lattice QCD state, highlights that the mass spectrum alone cannot resolve the nature of the $\Omega\Omega$ system and the decay properties of $\Omega\Omega$ dibaryon may provide an important criterion between a compact hexaquark and a weakly bound molecule.

The $\Omega\Omega$ dibaryon has not yet been observed experimentally, but both theoretical and experimental efforts suggest that it may be produced in high-energy heavy-ion collisions~\cite{Zhang:2020dma,Pu:2024kfh,Morita:2019rph,ExHIC:2017smd,Schaffner-Bielich:1999zoi,Xu:2002fw,Pal:2001rp,Shao:2004cn,Yu:2002jm}. Theoretically, the HAL QCD Collaboration has provided the $\Omega\Omega$ interaction potential from Lattice QCD~\cite{Gongyo:2017fjb}, and based on this potential, the production of  
$\Omega\Omega$ dibaryon and relevant tribaryons has been estimated within a dynamical coalescence mechanism for relativistic heavy-ion collisions at RHIC and LHC energies~\cite{Zhang:2020dma,Zhang:2021vsf}; more recent covariant coalescence studies show that the yield ratios $N_{\Omega\Omega}/N_{\Omega}$ exhibit a distinct centrality dependence between molecular and multiquark states, offering a promising way to probe the internal structure of the $\Omega\Omega$ dibaryon~\cite{Pu:2024kfh}. Experimentally, since the $\Omega\Omega$ system itself has not yet been reconstructed, femtoscopic measurements of $N\Omega$ correlations provide the most direct experimental access to this physics and offer valuable constraints for the closely related hyperon-hyperon interaction relevant to $\Omega\Omega$. The STAR Collaboration first measured the $p\Omega$ correlation function in Au+Au collisions, with the data favoring an $N\Omega$ bound state of about 27 MeV binding energy~\cite{STAR:2018uho}. The ALICE Collaboration subsequently confirmed the attractive interaction with high precision and tested Lattice QCD predictions~\cite{ALICE:2020mfd}. Moreover, the STAR, ALICE, and LHCb Collaborations are continuing their efforts to search for the $N\Omega$ bound state~\cite{Zhang:2025lfn}. These theoretical and experimental production studies are essential for assessing the feasibility of observing the $\Omega\Omega$ dibaryon. However, its detectability ultimately depends on its decay properties, which determine whether the state can be reconstructed in the detector and how its internal structure can be revealed. This provides the primary motivation for the systematic study of $\Omega\Omega$ weak decays undertaken in this work.

The weak decay properties of the $\Omega\Omega$ dibaryon remain almost entirely unexplored. Because the bound $\Omega\Omega$ state has strangeness $S=-6$, its strong decay is forbidden, and the only possible decay mechanism is the weak interaction, in which one or more strange quarks convert into light quarks, corresponding to the weak decay of one or both of the $\Omega$ baryons. The study of its decay is therefore not merely an accessory to its spectroscopy but also the only way to determine its lifetime and a crucial input for assessing the feasibility of experimental searches. Zhang et al. estimated its mean lifetime as approximately twice that of a free $\Omega$ hyperon of order $10^{-10}~{\rm s}$ with the chiral SU(3) quark model~\cite{Zhang:2000sv}. In terms of the quark mass density- and temperature-dependent model, Zhang and Su estimated the lifetime of the \((\Omega\Omega)_{0^+}\) state, finding it to be metastable at zero temperature with a lifetime of order \(10^{-11}~{\rm s}\), while the strong decay channel opens at temperatures around \(129~{\rm MeV}\)~\cite{Zhang:2003ka}. Separately, Schaffner-Bielich et al. developed a general SU(3) framework for the weak decays of strange dibaryons~\cite{Schaffner-Bielich:1999zoi}, but their analysis was restricted to two-body final states and therefore cannot be applied to the $\Omega\Omega$ system, where dominant decay modes of the $\Omega\Omega$ system are expected to be the three-body or even four-body final state. Since the decay width in a possible decay channel can be precisely measured in the experiment by reconstructing the initial particle from the tracks of decay products detected in a detector, and consequently the result might be used to distinguish the internal structure of the initial particle, compact hexaquark or weakly bound molecule, a systematic study of $\Omega\Omega$ weak decays, covering both three- and four-body channels, is both necessary and overdue.

In this work, we perform a systematic study of the weak decays and lifetime of the $\Omega\Omega$ dibaryon using an effective Lagrangian approach. We obtain the partial decay widths for the three- and four-body modes, and consequently derive the total weak decay width and the lifetime of the $\Omega\Omega$ dibaryon. Our results provide the theoretical reference values needed to assess the feasibility of experimental searches and offer a diagnostic tool for distinguishing between two different internal structures of the $\Omega\Omega$ dibaryon, namely a compact hexaquark and a weakly bound hadronic molecule. This work thus bridges the gap between the extensive efforts devoted to the mass and binding energy of the $\Omega\Omega$ dibaryon and the equally important but comparatively neglected question of how this exotic state decays.

The remainder of this paper is organized as follows. In Sec.~\ref{formalism}, we introduce the effective Lagrangian framework for the 
$\Omega\Omega$ dibaryon, present the decay amplitudes for the relevant decay channels, and include the three- and four-body phase space. In Sec.~\ref{results}, we perform the numerical results and discussion, including the decay widths and lifetime, as well as their implications for experimental searches and for discriminating between the compact hexaquark and loosely bound molecular interpretations. Finally, in Sec.~\ref{summary}, we summarize our findings and provide an outlook.

\section{Formalism}\label{formalism}

In the present work, we assume that the $J^P=0^+$ $\Omega\Omega$ dibaryon, denoted as $\mathcal{D}_{\Omega\Omega}$, is a bound state below the $\Omega\Omega$ threshold according to previous theoretical studies. Inspired by the phenomenological Lagrangians used to describe the deuteron as a $pn$ dibaryon~\cite{Weinberg:1962hj,Ivanov:1995zp,Dong:2008mt}\footnote{We note a typographical error of the effective Lagrangian for deuteron in Eq.~(74) of Weinberg's paper~\cite{Weinberg:1962hj}; the correct expression should read \(\mathcal{L}_{\mathrm{eff}} = G_{pnd} \bar{\psi}_p \gamma_\mu \psi_n^c \varphi_d^\mu\).}, we construct an effective Lagrangian to describe the interaction between the \(\mathcal{D}_{\Omega\Omega}\) dibaryon and two $\Omega$ baryons as
\begin{align}
	\mathcal{L}_{\mathcal{D}_{\Omega\Omega}\Omega\Omega}(x)
	&= \frac{1}{2}g_{\mathcal{D}_{\Omega\Omega}\Omega\Omega}
	\int dy\, \Phi(y^2)\, \mathcal{D}_{\Omega\Omega}(x)\,
	\bar{\Omega}^{C}_{\mu}(x+y/2) \nonumber \\
	&\quad \times \gamma_5 \Omega^{\mu}(x-y/2) + \mathrm{H.c.}.
\end{align}
Here, the notations $\psi^C=C \bar\psi^T$ and $\bar \psi^C= \psi^TC$ is adopted, where $C=i\gamma^2\gamma^0$ is the charge-conjugation matrix. The factor of $1/2$ serves as a symmetry factor for identical fields, ensuring that the effective Feynman vertex evaluates directly to $ig_{\mathcal{D}_{\Omega\Omega}\Omega\Omega}\gamma_5$ by canceling the factor of 2 arising from Wick contractions. The correlation function $\Phi(y^{2})$ is responsible for describing the distribution of the constituents in the relevant composite systems and rendering the Feynman diagrams ultraviolet finite. This correlation function $\Phi(y^{2})$ can be related to the its Fourier transform in momentum space $\tilde \Phi(-p^2)$ as 
\begin{align}
	\Phi(y^{2})
	&= \int \frac{d^4p}{(2\pi)^4}\, e^{-ipy}\, \tilde{\Phi}(-p^2) 
\end{align}
where $p$ is the 4-momentum of relative motion between the constituents in the Minkowski space. A Gaussian-like form of correlation function $\tilde \Phi(-p^2)$ is usually adopted, which can be expressed as~\cite{Dong:2017gaw,Dong:2008mt,Lu:2016nnt,Dong:2017rmg,Lin:2019qiv,Xiao:2020alj,Ling:2021asz}
\begin{equation}
 \tilde{\Phi}(-p^2)=\exp\left(p^2/\Lambda^2\right), \label{eq:phi}
\end{equation}
where the $\Lambda$ is the size parameter that has a dimension of
mass and defines a scale for the distribution of the constituents inside the composite particle.

\begin{figure}
	\centering
	\includegraphics[scale=0.50]{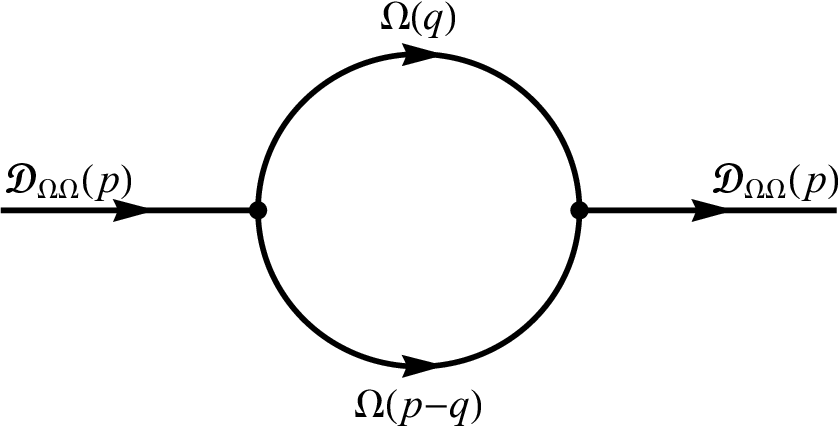}
	\caption{The Feynman diagram for the mass operator of $\mathcal{D}_{\Omega\Omega}$ dibaryon.}
	\label{mass}
\end{figure}

To calculate the couplings between the \(\mathcal{D}_{\Omega\Omega}\) dibaryon and its constituents, we employ the Weinberg-Salam compositeness condition, which states that the probability of finding the dressed bound state as a structureless state is equal to zero~\cite{Weinberg:1962hj,Salam:1962ap}. For the \(\mathcal{D}_{\Omega\Omega}\) state with \(J^P=0^+\), the compositeness condition reads
\begin{equation}
	Z_{\mathcal{D}_{\Omega\Omega}}=1-\left.\frac{\partial \Sigma(p^2)}{\partial p^2}\right|_{p^2=m_{\mathcal{D}_{\Omega\Omega}}^2}=0,\label{eq:com}
\end{equation}
where \(Z_{\mathcal{D}_{\Omega\Omega}}\) is the corresponding wave function renormalization constant, and \(\Sigma(p^2)\) is the self-energy of the dibaryon as shown in Fig.~\ref{mass}. This condition gives a direct relation between the coupling constant of the bound state to its constituents and its mass. The expression for the mass operator is given by 
\begin{align}
	\Sigma(p^2)
	&= -\frac{1}{2}(ig_{\mathcal{D}_{\Omega\Omega}\Omega\Omega})^2
	\int \frac{d^{4}q}{(2\pi)^{4}}\, \tilde{\Phi}^2\!\left[-(q-p/2)^2\right] \nonumber \\
	&\quad \times \mathrm{Tr}\!\left[ \gamma_5
	S^{\mu\nu}(q,m_{\Omega})\, \gamma_5 S^{c}_{\nu\mu}(p-q,m_{\Omega})
	\right], \label{eq:mas}
\end{align}
where \(S_{\mu\nu}(q,m)\) is the free fermion propagator for the spin-3/2 particle,
\begin{align}
	S_{\mu\nu}(q,m)
	&= i(\slashed{q}-m)^{-1} \nonumber \\
	&\quad \times \left(-g_{\mu\nu}
	+ \frac{\gamma_{\mu}\gamma_{\nu}}{3}
	+ \frac{2q_{\mu}q_{\nu}}{3m^{2}}
	+ \frac{\gamma_{\mu}q_{\nu}-\gamma_{\nu}q_{\mu}}{3m}
	\right),
\end{align}
and \(S_{\nu\mu}^c(q,m)\) is the corresponding charge-conjugate propagator,
\begin{align}
	S_{\nu\mu}^c(q,m) \equiv C S_{\mu\nu}^T(q,m) C^{-1} = S_{\nu\mu}(-q,m).
\end{align}
The $1/2$ in the mass operator $\Sigma(p^2)$ originates from the topological symmetry factor due to the exchange of identical internal propagators in the self-energy loop diagram.

Moreover, in the near-threshold nonrelativistic limit, by neglecting the finite-size effects, the compositeness condition yields a direct relation between the coupling constant and the binding energy, independent of the cutoff. It can be expressed as~\cite{Weinberg:1962hj}
\begin{equation}
	g_{\mathcal{D}_{\Omega\Omega}\Omega\Omega}^{NR}=4 \sqrt{\pi}\left(\frac{E_b}{m_{\Omega}}\right)^{1/4},
\end{equation}
where \(E_b = 2m_{\Omega}-m_{\mathcal{D}_{\Omega\Omega}}\) is the binding energy of the \(\mathcal{D}_{\Omega\Omega}\) dibaryon. In the near-threshold limit, this nonrelativistic result $g_{\mathcal{D}_{\Omega\Omega}\Omega\Omega}^{NR}$ serves as a reliable reference to calibrate the coupling constant $g_{\mathcal{D}_{\Omega\Omega}\Omega\Omega}$ obtained from our rigorous field-theoretical calculation. For larger binding energies, the nonrelativistic approximation breaks down, and the full covariant treatment with the form factor and cutoff becomes essential to accurately capture the dynamics of the compact bound state.

\begin{figure}
	\centering
	\includegraphics[scale=0.40]{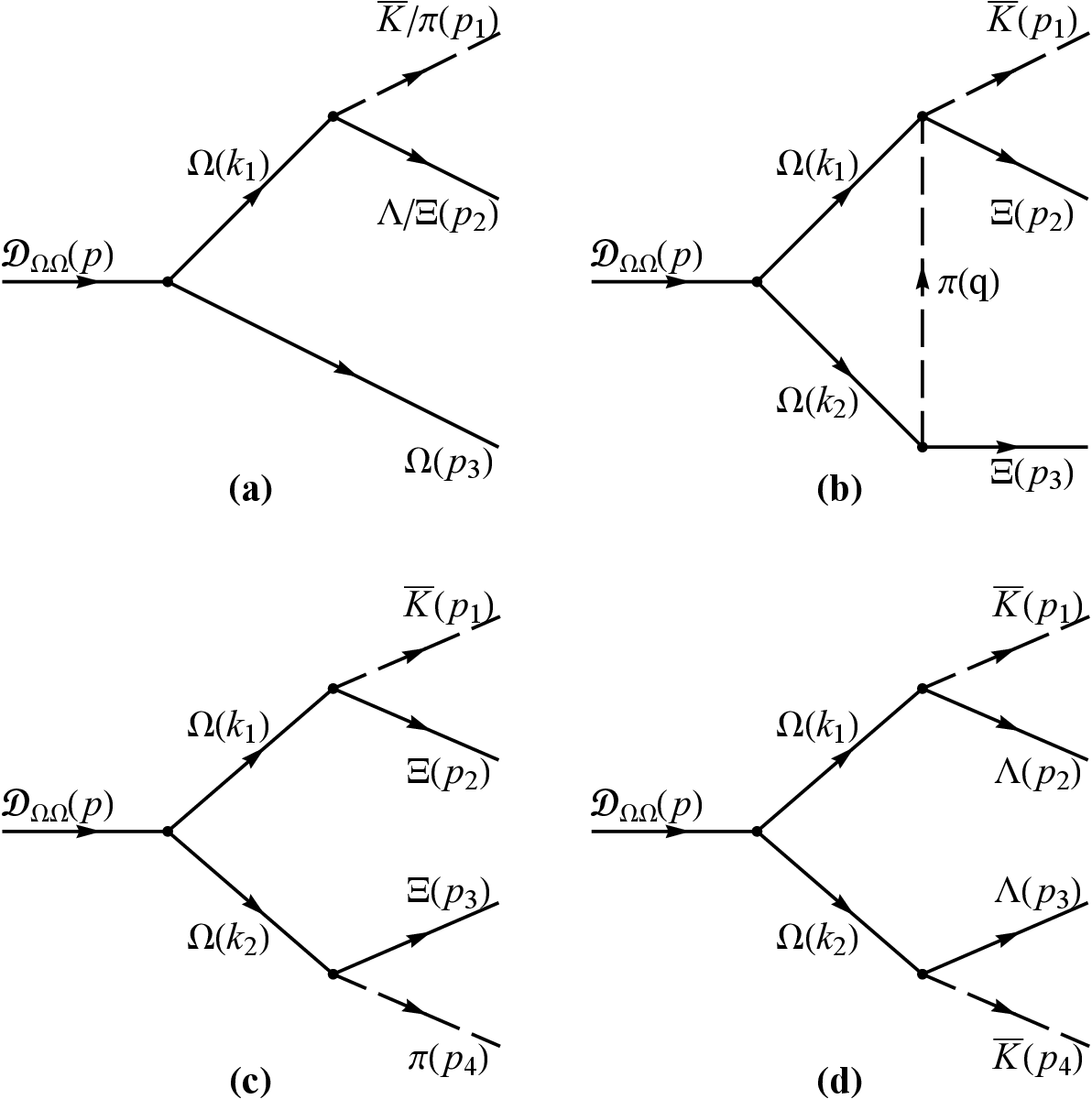}
	\caption{The Feynman diagrams for the weak decay modes of $\mathcal{D}_{\Omega\Omega}$ dibaryon. All direct diagrams are displayed. The additional diagrams arising from the exchange of identical final-state particles are not shown but are included in the calculations.}
	\label{feyn}
\end{figure}

The Feynman diagrams for the weak decays of the \(\mathcal{D}_{\Omega\Omega}\) dibaryon are shown in Fig.~\ref{feyn}. Two classes of diagrams are considered. The first class contains a single weak decay vertex, in which one of the two \(\Omega\) baryons decays weakly while the other either acts as a spectator or participates through a strong interaction vertex. Depending on the decay channel, this class can lead to either three-body or four-body final states. More explicitly, the weak decay vertices of the \(\Omega\) include \(\Omega \Lambda \bar{K}\), \(\Omega \Xi^0 \pi^-\), and \(\Omega \Xi^- \pi^0\). The isospin-breaking treatment of the \(\Omega \Xi^0 \pi^-\) and \(\Omega \Xi^- \pi^0\) interactions is necessary because the experimentally measured branching ratios of the free \(\Omega\) decays into \(\Xi^0 \pi^-\) and \(\Xi^- \pi^0\) differ significantly. The second class contains two weak decay vertices, in which both \(\Omega\) baryons decay weakly. In these channels, each weak decay produces two particles, so this class necessarily leads to four-body decay mode \(\Lambda \bar{K} \Lambda \bar{K}\).

If the \(\mathcal{D}_{\Omega\Omega}\) state is weakly bound, all the single weak vertex diagrams are kinematically allowed, and their contributions essentially saturate the total decay width. If the \(\mathcal{D}_{\Omega\Omega}\) state is deeply bound, the mass of the initial state is reduced, so that the \(\Omega \Lambda \bar{K}\) channel is forbidden by phase space. In this case, we consider the double weak vertex diagram \(\Lambda \bar{K} \Lambda \bar{K}\) as supplementary decay channel. We will consider various binding energy scenarios and discuss the resulting decay modes and lifetimes in the following section.

The relevant effective Lagrangians are written as~\cite{Borasoy:1999ip,Tandean:1998ch,Cao:2026llx,Shen:2025xcq,Huang:2018wth,Gao:2010ve}

\begin{equation}
	\mathcal{L}_{\Omega \Lambda \bar{K}}
	= \frac{g_{\Omega \Lambda \bar{K}}^{(\mathrm{PC})}}{m_{K}}\,
	\bar{\Lambda}\, \partial^\mu \bar{K}\, \Omega_{\mu}
	+ \mathrm{H.c.},
\end{equation}

\begin{equation}
	\mathcal{L}_{\Omega \Xi^0 \pi^-}
	= \frac{g_{\Omega \Xi^0 \pi^-}^{(\mathrm{PC})}}{m_{\pi^-}}\,
	\bar{\Xi}^0\, \partial^\mu \pi^-\, \Omega_{\mu}
	+ \mathrm{H.c.},
\end{equation}

\begin{equation}
	\mathcal{L}_{\Omega \Xi^- \pi^0}
	= \frac{g_{\Omega \Xi^- \pi^0}^{(\mathrm{PC})}}{m_{\pi^0}}\,
	\bar{\Xi}^-\, \partial^\mu \pi^0\, \Omega_{\mu}
	+ \mathrm{H.c.},
\end{equation}

\begin{equation}
	\mathcal{L}_{\Omega \Xi \bar K}=\frac{g_{\Omega \Xi \bar K}}{m_{K}}\bar{\Xi}\partial^\mu \bar K \Omega_{\mu}+{\rm H.c.},
\end{equation}

\begin{equation}
	\mathcal{L}_{\Omega \Xi \bar K^*}
	= i\, \frac{g_{\Omega \Xi \bar K^* }}{m_{K^*}}\,
	\bar{\Omega}^{\mu} \gamma^{\nu} \gamma_5
	\left[ \partial_{\mu} \bar K^*_{\nu} - \partial_{\nu} \bar K^*_{\mu} \right]
	\Xi + \mathrm{H.c.},
\end{equation}

\begin{equation}
	\mathcal{L}_{K^* K \pi}
	= i\, g_{K^* K \pi}\,
	\bar{K}^{*}_{\mu}
	\left(
	\vec{\pi} \cdot \vec{\tau}\, \partial^{\mu} K
	- \partial^{\mu} \vec{\pi} \cdot \vec{\tau}\, K
	\right).
\end{equation}
For the weak decay vertices, we adopt parity-conserving $P-$wave couplings and neglect the parity-violating $D-$wave contributions, which are strongly suppressed theoretically and not required by current experimental data on $\Omega$ decays~\cite{Borasoy:1999ip,ParticleDataGroup:2026mpi}. The four-point interaction $\mathcal{L}_{\Omega \pi \Xi \bar K}$ are induced by the exchange of a $K^*$ meson through the $\pi \Omega \to \bar K \Xi$ process. That is, with the effective Lagrangians $\mathcal{L}_{\Omega \Xi \bar K^*}$ and $\mathcal{L}_{K^* K \pi}$, we evaluate the corresponding amplitude of $\pi \Omega \to \bar K \Xi$ process via $t-$channel $K^*$ exchange and neglect the four momentum of $K^*$  to generate the contact interaction as
\begin{align}
	\mathcal{L}_{\Omega \pi \Xi \bar K}
	= & -\frac{2 g_{K^* K \pi} g_{\Omega \Xi K^*}}{m_{K^*}^3}\,
    \bar{\Xi}\,
\Bigl(
\partial^\nu K\,
\slashed{\partial} \vec{\pi} \cdot \vec{\tau}
- \partial^\nu \vec{\pi} \cdot \vec{\tau}\,
\slashed{\partial} K
\Bigr)
\gamma_5\, \Omega_\nu
\nonumber \\ & + \mathrm{H.c.}.
\end{align}
With the experimental partial decay widths~\cite{ParticleDataGroup:2026mpi}, we can obtain the coupling constants to be $g_{\Omega \Lambda \bar K}=1.98 \times 10^{-6}$, $g_{\Omega \Xi^0 \pi^-}=1.88 \times 10^{-7}$, $g_{\Omega \Xi^- \pi^0}=1.10 \times 10^{-7}$, and $g_{K^*K\pi}=-3.18$. Meanwhile, with the SU(3) symmetry, we determine the coupling constants to be $g_{\Omega \Xi \bar K} = 8.13$~\cite{Shen:2025xcq} and $g_{\Omega \Xi \bar K^*} = -7.01$~\cite{Huang:2018wth}.

With the effective Lagrangians constructed above, we now derive the decay amplitudes for the \(\mathcal{D}_{\Omega\Omega}\) dibaryon using standard Feynman rules. The four typical expressions shown in Fig.~\ref{feyn} read as follows

\begin{align}
	\mathcal{M}_{\Omega \Lambda \bar{K}}
	&= i g_{\mathcal{D}_{\Omega\Omega}\Omega\Omega}\,
	\frac{i g_{\Omega \Lambda \bar{K}}^{(\mathrm{PC})}}{m_{K}}\,
	\tilde{\Phi}\!\left[-(k_1-p/2)^2\right] i p_{1\mu} \nonumber \\
	&\quad \times \bar{u}(p_2,s_2)\,
	S^{\mu\nu}(k_1,m_{\Omega})\,
	\gamma_5 C \bar{u}^{T}_{\nu}(p_3,s_3),
\end{align}

\begin{align}
	\mathcal{M}_{\Xi^- \Xi^0 K^-}^{\pi^0}
	&= i g_{\mathcal{D}_{\Omega\Omega}\Omega\Omega}\,
	\frac{-2 i g_{K^* K \pi} g_{\Omega \Xi K^*}}{m_{K^*}^3}\,
	\frac{i g_{\Omega \Xi^- \pi^0}^{(\mathrm{PC})}}{m_{\pi^0}}\,
	 \int \frac{d^4q}{(2\pi)^4} \nonumber \\
	&\quad \times \tilde{\Phi}\!\left[-(k_1-p/2)^2\right]
	\bar{u}(p_2,s_2)\,
	\left(p_{1\mu} \slashed{q} - q_\mu \slashed{p}_1\right) \gamma_5 \nonumber \\
	&\quad \times S^{\mu\nu}(k_1,m_{\Omega})\,
	\gamma_5 S^{c}_{\nu \alpha}(k_2,m_{\Omega})\,
	C \bar{u}^T(p_3,s_3)\,
	i q^\alpha\, \nonumber \\
	&\quad \times\frac{i}{q^2-m_{\pi^0}^2},
\end{align}

\begin{align}
	\mathcal{M}_{\Xi^0 K^- \Xi^- \pi^0}
	&= i g_{\mathcal{D}_{\Omega\Omega}\Omega\Omega}\,
	\frac{i g_{\Omega \Xi \bar{K}}}{m_{K}}\,
	\frac{i g_{\Omega \Xi^- \pi^0}^{(\mathrm{PC})}}{m_{\pi^0}}\,
	\tilde{\Phi}\!\left[-(k_1-p/2)^2\right] \nonumber \\
	&\quad \times \bar{u}(p_2,s_2)\,
	S^{\mu\nu}(k_1,m_{\Omega})\,
    \gamma_5 S^{c}_{\nu \alpha}(k_2,m_{\Omega})\, \nonumber \\
	& \quad \times  C \bar{u}^T(p_3,s_3) i p_{1\mu} i p_4^\alpha  ,
\end{align}

\begin{align}
{\cal M}_{\Lambda \bar K \Lambda \bar K}
	&= i g_{\mathcal{D}_{\Omega\Omega}\Omega\Omega}\,
	 \frac{ig_{\Omega \Lambda \bar{K}}^{(\mathrm{PC})}}{m_{K}} \,
\frac{ig_{\Omega \Lambda \bar{K}}^{(\mathrm{PC})}}{m_{K}} \,
	\tilde{\Phi}\!\left[-(k_1-p/2)^2\right] \nonumber \\
	&\quad \times \bar{u}(p_2,s_2)\,
	S^{\mu\nu}(k_1,m_{\Omega})\,
	\gamma_5 S^{c}_{\nu \alpha}(k_2,m_{\Omega})\, \nonumber \\
	& \quad \times  C \bar{u}^T(p_3,s_3) i p_{1\mu} i p_4^\alpha + \text{exchange terms}.
\end{align}
For the three-body tree-level diagrams, the amplitudes for the \(\Omega \Xi^0 \pi^-\) and \(\Omega \Xi^- \pi^0\) decay modes can be obtained in a similar way as for \(\mathcal{M}_{\Omega \Lambda \bar{K}}\), with the corresponding couplings and fields replaced accordingly. For the loop diagrams, the resulting amplitudes \(\mathcal{M}_{\Xi^- \Xi^0 K^-}^{\pi^-}\) and \(\mathcal{M}_{\Xi^- \Xi^- \bar{K}^0}^{\pi^0}\) also share the same form as \(\mathcal{M}_{\Xi^- \Xi^0 K^-}^{\pi^0}\), where the superscript \(\pi^0\) or \(\pi^-\) denotes the exchanged meson in the loop. The total amplitude for the \(\Xi^- \Xi^0 K^-\) decay mode is the coherent sum of \(\mathcal{M}_{\Xi^- \Xi^0 K^-}^{\pi^0}\) and \(\mathcal{M}_{\Xi^- \Xi^0 K^-}^{\pi^-}\). The amplitudes for the various four-body decay modes \(\mathcal{M}_{\Xi \bar K \Xi \pi}\) can be obtained analogously to \(\mathcal{M}_{\Xi^0 K^- \Xi^- \pi^0}\), with the appropriate replacements of couplings and fields. 

For final states containing identical particles, such as \(\Xi^0 K^- \Xi^0 \pi^-\), \(\Xi^- \bar K^0 \Xi^- \pi^0\), \(\Xi^- \Xi^- \bar{K}^0\), and \(\Lambda\bar{K} \Lambda\bar{K}\), the exchange contributions must be included through the symmetrization procedure; that is, the amplitude is symmetrized for identical bosons and antisymmetrized for identical fermions, with an extra minus sign arising from each exchange of two identical fermions.

The differential partial decay width of the \(\mathcal{D}_{\Omega\Omega}\) dibaryon into an \(n\)-body final state can be expressed as
\begin{equation}
	d\Gamma_f = \frac{(2\pi)^4}{2m_{\mathcal{D}_{\Omega\Omega}}}\, \overline{|\mathcal{M}_f|^2}\, d\Phi_n,
\end{equation}
where the \(n\)-body phase space is given by
\begin{equation}
	d\Phi_n
	= \frac{1}{\prod_{\alpha} n_\alpha!}\, \delta^{(4)}\!\left(p - \sum_{i=1}^{n} p_i\right)
	\prod_{j=1}^{n} \frac{d^3 \boldsymbol{p}_j}{(2\pi)^3\, 2E_j}.
\end{equation}
Here, \(p\) and \(p_i\) are the four-momenta of the initial state and the \(i\)-th particle in the final state, respectively; $E_j$ is the energy of the of the \(j\)-th final-state particle. The \(\alpha\) labels the species of identical particles in the final state, \(n_\alpha\) is the number of particles of species \(\alpha\), and the factor \(1/\prod_\alpha n_\alpha!\) removes the overcounting of identical particles. The quantity \(\overline{|\mathcal{M}_f|^2}\) denotes the squared amplitude for channel $f$, averaged over the initial spins and summed over the final spins. General methods for dealing with the multi-body phase space can be found in Refs.~\cite{Jing:2020tth,Yu:2021dtp,Song:2022kac,Yang:2026bhm}. With the above amplitudes, the partial decay width for a given channel is obtained by integrating the differential partial decay width, and the total decay width is then obtained by summing over all partial decay widths,
\begin{equation}
	\Gamma_{\rm total}
	= \sum_f \Gamma_f
	= \sum_f \int d\Gamma_f.
\end{equation}
Finally, the lifetime of the \(\mathcal{D}_{\Omega\Omega}\) dibaryon is
\begin{equation}
	\tau
	= \frac{1}{\Gamma_{\rm total}}.
\end{equation}

\section{Results and discussion}\label{results}

\begin{figure}
	\centering
	\includegraphics[scale=0.52]{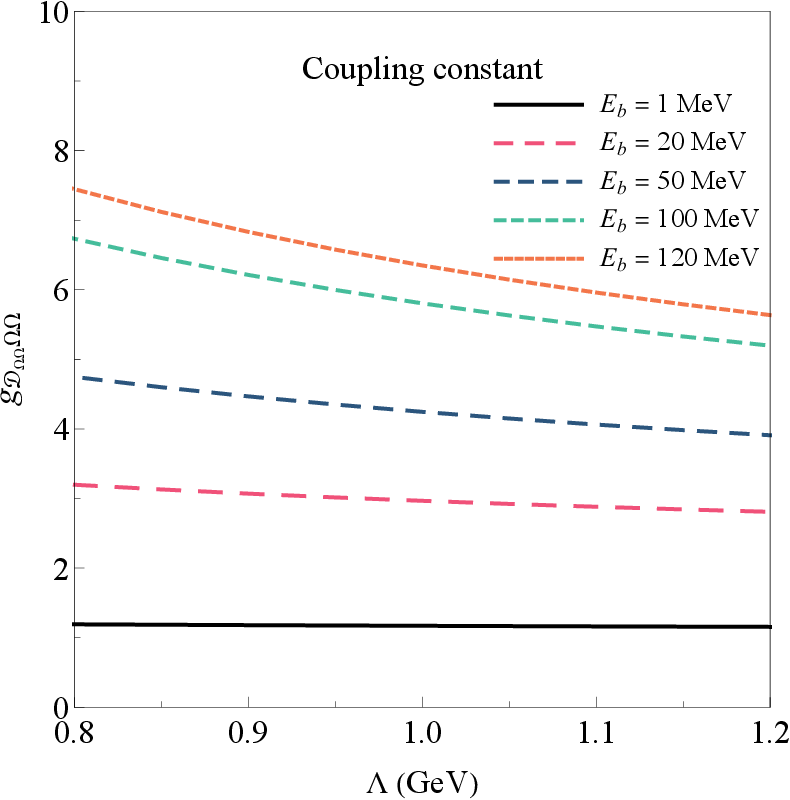}
	\caption{The coupling constant \(g_{\mathcal{D}_{\Omega\Omega}\Omega\Omega}\) as a function of the cutoff parameter \(\Lambda\) and binding energy \(E_b\).}
	\label{coupling}
\end{figure}

We present the coupling constant \(g_{\mathcal{D}_{\Omega\Omega}\Omega\Omega}\) as a function of the cutoff for several representative values of the binding energy: \(E_b=1\), \(20\), \(50\), \(100\), and \(120\) MeV. These values are chosen to cover the range from weak binding near threshold to deep binding, and are not intended to correspond precisely to any specific model prediction, given the discrepancies among different theoretical approaches as well as the intrinsic uncertainties within each approach. In particular, \(E_b = 120\) MeV is near the upper limit allowed by the chiral SU(3) quark model. This $E_b$ value is beyond the range predicted by several other approaches, and only serves to illustrate the behavior of the coupling in this limit. Following previous effective Lagrangian studies of hadronic molecules, where a cutoff around \(1~\mathrm{GeV}\) is typically employed~\cite{Chen:2015igx,Lu:2016nnt,Dong:2017rmg,Xiao:2019mvs,Lin:2019qiv,Guo:2019fdo,Faessler:2007gv,Dong:2009tg}, we vary \(\Lambda\) within the range \(0.8\)--\(1.2\) GeV to estimate the theoretical uncertainties in the decay observables. The predicted coupling constant \(g_{\mathcal{D}_{\Omega\Omega}\Omega\Omega}\) as a function of \(\Lambda\) is displayed in Figure~\ref{coupling}. It can be seen that the coupling constant \(g_{\mathcal{D}_{\Omega\Omega}\Omega\Omega}\) decreases slowly with increasing cutoff parameter \(\Lambda\), while it increases as the binding energy becomes larger. The resulting range of predictions for each fixed binding energy is presented in Table~\ref{coup}.

The relativistic coupling constant \(g_{\mathcal{D}_{\Omega\Omega}\Omega\Omega}\) exhibits a running behavior with respect to the cutoff parameter, but the variation is rather mild. This is expected, since the coupling constant for an \(S\)-wave state should not depend strongly on the cutoff. Meanwhile, the nonrelativistic result \(g_{\mathcal{D}_{\Omega\Omega}\Omega\Omega}^{NR}\) is also presented for comparison. It is observed that, owing to the inclusion of the form factor, the relativistic coupling constant is consistently larger than its nonrelativistic counterpart. For small binding energies, the two values are comparable, whereas for large binding energies the discrepancy becomes significant. This stems from the fact that the square of the momentum in Eq.~\eqref{eq:phi} is negative for a bound state. Consequently, such a form factor reduces the loop function $\Sigma(p^2)$ in Eqs.~\eqref{eq:com} and~\eqref{eq:mas}, resulting in a slightly larger solved coupling constant than that obtained without the form factor. Obviously, this effect becomes stronger for deeper bound states.

\begin{table}[!htbp]
	\centering
		\caption{The coupling constant \(g_{\mathcal{D}_{\Omega\Omega}\Omega\Omega}\) as a function of the binding energy \(E_b\). The range corresponds to varying the cutoff \(\Lambda\) from \(0.8\) to \(1.2~\mathrm{GeV}\), and the nonrelativistic result \(g_{\mathcal{D}_{\Omega\Omega}\Omega\Omega}^{NR}\) is also presented for comparison.}
	\label{coup}
	\begin{tabular*}{\columnwidth}{@{\extracolsep{\fill}}ccc}
		\hline\hline
		$E_b$ (MeV) & $g_{\mathcal{D}_{\Omega\Omega}\Omega\Omega}$ & \(g_{\mathcal{D}_{\Omega\Omega}\Omega\Omega}^{NR}\) \\
		\hline
		1   & 1.19--1.16 & 1.11\\
		20  & 3.20--2.81 & 2.34\\
		50  & 4.75--3.91 & 2.95\\
		100 & 6.73--5.20 & 3.51\\
		120 & 7.45--5.64 & 3.67\\
		\hline\hline
	\end{tabular*}
\end{table}

We first consider the weakly bound scenario with \(E_b = 1\) and \(20\) MeV, and the numerical results are listed in Table~\ref{decay1}. In this case, all the single-weak-vertex diagrams are kinematically allowed and exhaust essentially the entire total decay width. Also, the three-body channels \(\Omega \Lambda \bar{K}\), \(\Omega \Xi \pi\), and \(\Xi \Xi \bar{K}\) account for almost all of the width, while the four-body channel \(\Xi \bar{K} \Xi \pi\) is negligibly small. For \(E_b = 1\) MeV, \(\Omega \Lambda \bar{K}\) is the largest channel, while for \(E_b = 20\) MeV the \(\Omega \Xi \pi\) and \(\Xi \Xi \bar{K}\) channels also become significant. It is worth noting that the largest channel shifts from \(\Omega \Lambda \bar{K}\) at \(E_b=1\) MeV to \(\Omega \Xi \pi\) at \(E_b=20\) MeV due to the differing phase space suppressions. Also, the \(\Omega \Xi^0 \pi^-\) and \(\Omega \Xi^- \pi^0\) channels exhibit distinct partial widths, reflecting the isospin-breaking effects incorporated in our calculation. Regarding the cutoff dependence, the partial widths of the dominant tree-level channels, such as \(\Omega \Lambda \bar{K}\) and \(\Omega \Xi \pi\), decrease mildly with increasing \(\Lambda\), reflecting the slow decrease of the effective coupling \(g_{\mathcal{D}_{\Omega\Omega}\Omega\Omega}\) with \(\Lambda\). Meanwhile, the partial width of the \(\Xi \Xi \bar{K}\) channel, which is generated via loop and rescattering mechanisms, exhibits a significant increase with larger \(\Lambda\) as high-momentum contributions are included. The resulting lifetime is \(\tau = (0.51\text{--}1.62)\times 10^{-10}~\mathrm{s}\), corresponding to a proper decay length of \(c\tau = 1.53\text{--}4.86~\mathrm{cm}\), which is long enough for searching the \(\Omega\Omega\) dibaryon experimentally.

\begin{table*}[!htbp]
	\centering
	\small
	\caption{Decay widths and lifetimes for \(E_b = 1\), 20, and 50 MeV, with the cutoff \(\Lambda\) varying from \(0.8\) to \(1.2\)~GeV. The numbers following each decay mode denote the thresholds in MeV. For the total width and lifetime columns, the values in parentheses indicate the local extremum (minimum for total width, maximum for lifetime) within the scanned \(\Lambda\) range, while the outer values correspond to the endpoints.}
	\label{decay1}
	\begin{tabular*}{\textwidth}{@{\extracolsep{\fill}}cccccccc}
		\hline\hline
		\(E_b\) (MeV) & \multicolumn{5}{c}{\(\Gamma_i\) (\(10^{-13}\)~MeV)} & \(\Gamma_{\rm total}\) (\(10^{-13}\)~MeV) & \(\tau\) (\(10^{-10}\)~s) \\
		\cline{2-6}
		& \(\Omega \Lambda \bar{K}\) (3282) & \(\Omega \Xi^0 \pi^-\) (3127) & \(\Omega \Xi^- \pi^0\) (3129) & \(\Xi \Xi \bar{K}\) (3130) & \(\Xi \bar{K} \Xi \pi\) (3263) & & \\
		\hline
		1   & 81.43--77.61 & 34.23--32.98 & 12.03--11.59 & 0.62--2.10 & \((8.85\text{--}8.24)\times 10^{-5}\) & 128.31--124.28 & 0.51--0.53 \\
		20  & 15.77--12.57 & 19.31--16.21 & 6.73--5.65 & 1.76--6.77 & \((14.73\text{--}11.13)\times 10^{-5}\) & 43.57--(40.68)--41.20 & 1.51--(1.62)--1.60 \\
		50  & 0.36--0.24  & 10.15--7.60  & 3.50--2.62   & 1.48--6.12 & \((1.11\text{--}0.73)\times 10^{-5}\) & 15.49--(14.89)--16.58  & 4.25--(4.42)--3.97 \\
		\hline\hline
	\end{tabular*}
\end{table*}

For the intermediate binding energy \(E_b = 50\) MeV, the decay pattern remains qualitatively similar to the weakly bound case. The three-body channels still dominate the total width, while the four-body contributions remain negligible. Moreover, the \(\Omega \Lambda \bar{K}\) channel is strongly suppressed due to the reduced phase space, and the \(\Omega \Xi \pi\) and \(\Xi \Xi \bar{K}\) channels dominate. For \(E_b = 50\) MeV, although the coupling constant increases about  (237\text{--}299)\% and (39\text{--}48)\% than those of the \(E_b = 1\) and \(20\) MeV cases respectively,  the greatly reduced phase space leads to a smaller total width than at \(E_b = 1\) and \(20\) MeV. Then, the lifetime and proper decay length are correspondingly longer, \(\tau = (3.97\text{--}4.42)\times 10^{-10}~\mathrm{s}\) and \(c\tau = 11.90\text{--}13.25~\mathrm{cm}\), as listed in Table~\ref{decay1}.

In the deeply bound scenario with \(E_b = 100\) and \(120\) MeV, the mass of the initial dibaryon is reduced, so that the three-body \(\Omega \Lambda \bar{K}\) and four-body  \(\Xi \bar{K} \Xi \pi\) channels are forbidden by phase space. The numerical results are displayed in Table~\ref{decay2}. The dominant decay modes are the three-body \(\Omega \Xi \pi\) and \(\Xi \Xi \bar{K}\) channels. As discussed in the weakly bound case, the isospin-breaking effects lead to distinct partial widths for the \(\Omega \Xi^0 \pi^-\) and \(\Omega \Xi^- \pi^0\) channels owing to the independent effective couplings extracted from the experimentally measured branching ratios of free $\Omega$ decays. The supplementary double-weak-vertex channel \(\Lambda \bar{K} \Lambda \bar{K}\) is found to be negligibly small. With the phase space reduced and some channels forbidden, the total width is significantly smaller than in the weakly bound case. The resulting lifetime is \(\tau = (12.21\text{--}26.43)\times 10^{-10}~\mathrm{s}\), corresponding to a proper decay length of \(c\tau = 36.61\text{--}79.24~\mathrm{cm}\), approximately one order of magnitude longer than in the weakly bound scenario. Such a long lifetime means that the \({\Omega\Omega}\) dibaryon may also be searched for experimentally, although the search strategies may differ from those for short-lived states.

\begin{table*}[!htbp]
	\centering
	\small
	\caption{The notation is the same as in Table~\ref{decay1}, but for \(E_b = 100\) and 120 MeV. With the \(\Omega \Lambda \bar{K}\) and \(\Xi \bar{K} \Xi \pi\) channels forbidden by phase space, the \(\Lambda \bar{K} \Lambda \bar{K}\) mode is included.}
	\label{decay2}
	\begin{tabular*}{\textwidth}{@{\extracolsep{\fill}}ccccccc}
		\hline\hline
		\(E_b\) (MeV) & \multicolumn{4}{c}{\(\Gamma_i\) (\(10^{-13}\)~MeV)} & \(\Gamma_{\rm total}\) (\(10^{-13}\)~MeV) & \(\tau\) (\(10^{-10}\)~s) \\
		\cline{2-5}
		& \(\Omega \Xi^0 \pi^-\) (3127) & \(\Omega \Xi^- \pi^0\) (3129) & \(\Xi \Xi \bar{K}\) (3130) & \(\Lambda \bar{K} \Lambda \bar{K}\) (3219) & & \\
		\hline
		100 & 3.13--2.03 & 1.06--0.68 & 0.58--2.68 & \((9.55\text{--}5.78)\times 10^{-16}\) & 4.77--(4.52)--5.39 & 13.80--(14.56)--12.21 \\
		120 & 1.74--1.07 & 0.58--0.36 & 0.34--1.59 & \((6.43\text{--}3.70)\times 10^{-19}\) & 2.66--(2.49)--3.02 & 24.74--(26.43)--21.79 \\
		\hline\hline
	\end{tabular*}
\end{table*}

\begin{figure*}
	\centering
	\includegraphics[scale=0.52]{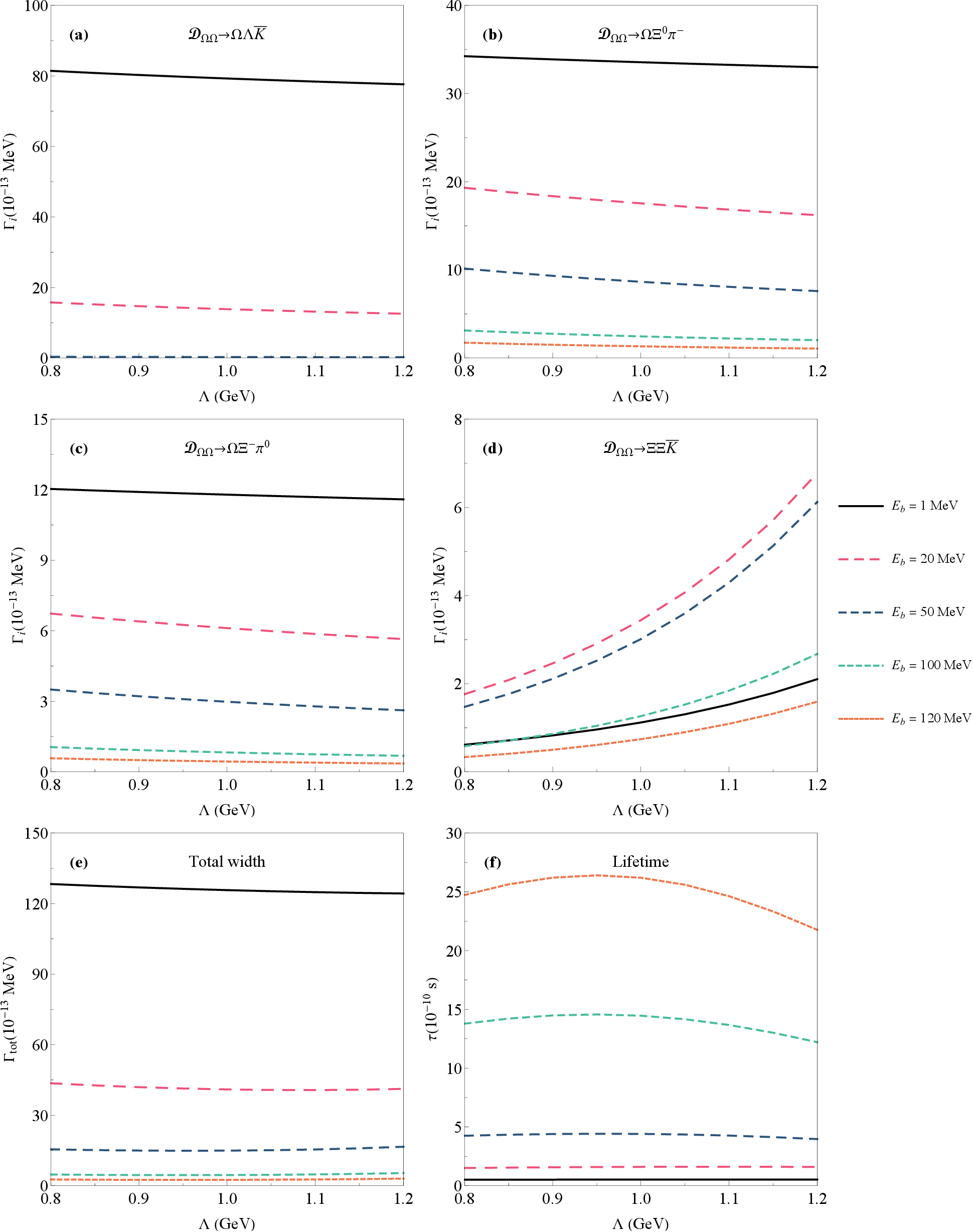}
	\caption{The partial decay widths, total width, and lifetime as functions of the cutoff parameter \(\Lambda\) and binding energy \(E_b\). The four-body channels \(\Xi \bar{K} \Xi \pi\) and \(\Lambda \bar{K} \Lambda \bar{K}\) are omitted, as their widths are negligibly small.}
	\label{width}
\end{figure*}

To systematically visualize the competing mechanisms across different binding energies, we display the dependence of the partial widths, total width, and lifetime on the cutoff parameter \(\Lambda\) in Fig.~\ref{width}. It can be seen that the tree-level channels \(\Omega \Lambda \bar{K}\) and \(\Omega \Xi \pi\)) decrease mildly with increasing \(\Lambda\), while the loop-induced \(\Xi \Xi \bar{K}\) channel increases significantly at larger \(\Lambda\). The relatively rapid increase of the \(\Xi \Xi \bar{K}\) channel originates from the fact that a larger cutoff opens up the high-momentum region of the loop integral, thereby releasing the short-distance contributions necessary for hadron-hadron rescattering. This enhancement significantly overwhelms the suppression caused by the decreasing effective coupling constant, which is a general feature of such triangle loop diagrams. Due to the interplay between the decreasing tree-level channels and the increasing loop-induced \(\Xi \Xi \bar{K}\) channel, except for the monotonic decrease at \(E_b = 1\) MeV,  the total decay width shows a non-monotonic behavior for intermediate and deeply bound states, exhibiting a local minimum within the scanned interval. Correspondingly, the lifetime reaches a local maximum with the cutoff $\Lambda$ varying from 0.8 to 1.2 GeV. This non-trivial dependence on the cutoff underscores the importance of treating both tree-level and loop-induced diagrams consistently, and demonstrates how the interplay between them shapes the observable decay properties.

Having clarified the cutoff dependence, we now compare the different binding energy regimes, where a clear qualitative difference in the decay patterns emerges. For weakly bound cases (\(E_b = 1\) and \(20\) MeV), the three-body channels \(\Omega \Lambda \bar{K}\), \(\Omega \Xi \pi\), and \(\Xi \Xi \bar{K}\) dominate. For deeply bound cases (\(E_b = 100\) and \(120\) MeV), the \(\Omega \Lambda \bar{K}\) channel is strictly forbidden by phase space, leaving the \(\Omega \Xi \pi\) and \(\Xi \Xi \bar{K}\) channels as the sole dominant contributions. The \(\Xi \Xi \bar{K}\) channel exhibits a non-monotonic dependence on \(E_b\), peaking around \(20\) MeV and then decreasing for deeply bound states. This behavior originates from the combined effects of the available phase space, the effective coupling constant, and the four-momentum dependence in the loop amplitude. On the other hand, the four-body processes remain negligibly small across all binding energies. Specifically, the \(\Xi \bar{K} \Xi \pi\) decay mode arising from the single-weak-vertex diagram is highly suppressed by phase space, while the \(\Lambda \bar{K} \Lambda \bar{K}\) channel is further suppressed by the additional weak interaction. As a physical consistency check, we note that at the weak binding limit (\(E_b = 1\) MeV), the calculated total width is slightly smaller than \(2\Gamma_\Omega\), which aligns perfectly with the physical picture of a loosely bound molecular state where the two \(\Omega\) baryons decay almost independently. As the binding energy increases, the total width deviates further from this limit and becomes significantly smaller, indicating a more compact internal structure. The lifetime increases from the order of \(10^{-10}~\mathrm{s}\) at weak binding to the order of \(10^{-9}~\mathrm{s}\) at deep binding. This distinct difference in the dominant decay modes and in the lifetime provides a useful discriminator between the compact hexaquark and loosely bound molecular interpretations of the \(\Omega\Omega\) dibaryon.

Experimentally, the \(\Omega\Omega\) dibaryon can be searched for using data from the STAR Collaboration at RHIC and the ALICE and LHCb Collaborations at the LHC. The closely related \(N\Omega\) system has been widely studied both theoretically and experimentally, demonstrating the feasibility of searching for strangeness dibaryons in heavy-ion collisions. In contrast, theoretical studies of the \(\Omega\Omega\) dibaryon have so far focused mainly on its spectrum and production, leaving its decay properties almost unexplored. Our calculations of its decay modes and lifetimes across different binding energies provide the theoretical reference values needed to guide these experimental searches and to probe the internal structure of this exotic state.

\section{Summary}\label{summary}

In this work, we perform a systematic study of the weak decays and lifetime of the \(\Omega\Omega\) dibaryon using an effective Lagrangian approach. The coupling constant \(g_{\mathcal{D}_{\Omega\Omega}\Omega\Omega}\) is determined from the compositeness condition, and its dependence on the binding energy \(E_b\) and the cutoff \(\Lambda\) is taken into account by varying \(\Lambda\) in the range \(0.8\)--\(1.2\) GeV for each representative value of \(E_b\). We derive the decay amplitudes for both single- and double-weak-vertex diagrams, and calculate the partial widths for the three- and four-body final states by integrating the squared amplitudes over the corresponding phase space. The total decay widths and lifetimes for different binding energies are then obtained as the main results of this work.

For weakly bound states with \(E_b = 1\) and \(20\) MeV, all the single-weak-vertex diagrams are kinematically allowed, and the three-body channels \(\Omega \Lambda \bar{K}\), \(\Omega \Xi \pi\), and \(\Xi \Xi \bar{K}\) exhaust essentially the entire width. With the increase of binding energy, the largest contribution shifts from \(\Omega \Lambda \bar{K}\) at \(E_b = 1\) MeV to \(\Omega \Xi \pi\) at \(E_b = 20\) MeV, while \(\Xi \Xi \bar{K}\) also provides a significant contribution. The lifetime is of order \(10^{-10}~\mathrm{s}\) for these weakly bound cases. The intermediate case \(E_b = 50\) MeV shows a similar pattern, although the \(\Omega \Lambda \bar{K}\) channel is strongly suppressed by phase space. For deeply bound states with \(E_b = 100\) and \(120\) MeV, the \(\Omega \Lambda \bar{K}\) channel is forbidden, and the dominant decay modes are the three-body \(\Omega \Xi \pi\) and \(\Xi \Xi \bar{K}\) channels. The supplementary double-weak-vertex channel \(\Lambda \bar{K} \Lambda \bar{K}\) is found to be negligibly small. The total width is significantly smaller, and the lifetime increases to the order of \(10^{-9}~\mathrm{s}\), approximately one order of magnitude longer than in the weakly bound cases. This difference in decay modes and lifetimes provides a clear discriminator between the compact hexaquark and loosely bound molecular interpretations of the \(\Omega\Omega\) dibaryon.

The \(\Omega\Omega\) dibaryon can be searched for using data from the STAR Collaboration at RHIC and the ALICE and LHCb Collaborations at the LHC. It can be probed through femtoscopic correlation measurements, as well as through the invariant-mass reconstruction of its decay products. Our calculations of the decay modes and lifetimes for different binding energies provide theoretical reference values for these experimental searches.

\subsection*{ACKNOWLEDGMENTS}

We would like to thank Li-Sheng Geng for helpful discussions. This work is supported by the Scientific Research Foundation of Hunan Provincial Education Department under Grant No. 24B0063, and the Youth Talent Support Program of Hunan Normal University under Grant No. 2024QNTJ14. This work is supported by the National Natural Science Foundation of China under Grants No. 12375142, No. 12547111, and No. 12221005, and by the Chinese Academy of Sciences under Grant No. YSBR-101.

\end{document}